%% file: Escapade.tex
\documentclass[showpacs,preprintnumbers,amsmath,amssymb,aps,twocolumn]{revtex4}

\usepackage{graphicx}
\usepackage{amssymb}
\usepackage{bm}%
\usepackage{hyperref}

\input{commands.tex}

\begin{document}

\title{Cellular Computing and Least Squares for partial differential problems parallel solving}

\author{N.Fressengeas}
\AffilMOPS

\author{H.Frezza-Buet}
\AffilIMS



%
%
%


\begin{abstract}
 This paper shows how partial differential problems can be solved thanks to cellular computing and an adaptation of the Least Squares Finite Elements Method. As cellular computing can be implemented on distributed parallel architectures, this method allows the distribution of a resource demanding differential problem over a computer network.
\end{abstract}


%



\maketitle

\section{Partial differential equations and cellular computing}
\subsection{Automata and calculus}
Ever since von Neumann~\cite{neumann}, the question of modeling
continuous physics with a discrete set of cellular automata
has been raised, whether they handle discrete or continuous values.
Many answers have been brought forth
through, for instance, the work of Stephen
Wolfram~\cite{wol83rmp601} summarized in a recent
book~\cite{wolfram}. This problem has been mostly tackled by rightfully
considering that modeling physics through Newton and Leibniz
calculus is fundamentally different from a discrete modelisation as
implied by automata.

Indeed, the former implies that physics is considered continuous
either because materials and fields are considered continuous in
classical physics or because quantum physics wave functions are
themselves continuous. On the contrary, modeling physics through
automata implies modeling on a discrete basis, in which a unit
element called \emph{a cell}, interacts with its surroundings according to  a
given law derived from local physics considerations.

Such discrete
automaton based models have been successfully applied to various
applications ranging from reaction-diffusion
systems~\cite{wei94prl1749} to forest fires~\cite{dro92prl1629}, through
probably one of the most impressive achievements: the Lattice Gas
Automata~\cite{rot94rmp1417}, where atoms or molecules are considered
individually. 
In this frame, simple point mechanics interaction rules lead to
complex behaviors such as phase transition and turbulence. This
peculiar feature of automata, making complex group behavior emerge
from fairly simple individual rules aroused the interest around them
for the past decades \cite{Chopard98}.

\subsection{Cellular computing}

Cellular automata-based modeling attempts have also concerned the theory of circuits
for a few decades, because the Very-Large-Scale Integration (VLSI) components offer a large amount
of configurable processors, spatially organized as a locally connected
array of analogical and numerical processing units. In this field, the
concept of cellular automata can be extended \cite{chua88} by allowing local cells,
that are dynamical systems, to deal with several continuous values and
local connections.

Such cellular computing algorithms are good candidates for the numerical
resolution of partial differential equations (PDE), and a
methodology for their design from a given PDE has been proposed
in~\cite{roska95}. This approach consists in performing a spatial
discretisation of the PDE through the finite difference
scheme, yielding an Ordinary Differential Equation (ODE) on time that can be numerically solved by standard methods like Runge-Kutta.

This approach is widely used in this field, and drives the design of simulators like SCNN
2000~\cite{loncar00}, as well as the design of actual VLSI components~\cite{sargeni05}. The partial differential system is there implemented using analogical VLSI components, the circuit temporal evolution being then the temporal evolution of the initial PDE.

Two main difficulties arise in this framework. The first one concerns the stability of the cellular system. Some stability studies of cellular networks for classical PDEs can be found in~\cite{slavova00} but stability has still to be analyzed when dealing with new specific problems, as it has been done, for example, for the dynamics of nuclear reactors~\cite{hadad07}.
The second difficulty raised by transforming PDE to ODE for resolution by cellular means is the actual fitting of the cellular algorithm to the PDE, since the method is more a heuristic one than a formal derivation from the PDE, as mentioned in~\cite{bandmann02}. Furthermore, the features of the cellular algorithm cannot be easily associated to the physical parameters involved in the PDE.

To cope with the lack of methods to formally derive a cellular algorithm from a PDE, some parameter tunings can be performed. This tuning can be driven by a supervised learning process, as in~\cite{loncar00,bandmann02}. Some other a posteriori checks can be achieved if some analytical solution of the PDE is known for particular cases, as in~\cite{slavova00}, or if some behavior  can be expected, as traveling waves or solitons~\cite{roska95,kozek95,slavova03}. In the latter case, validation is based on a qualitative criterion.

Some other methods to derive automata from particular differential problems such as reaction-diffusion systems~\cite{wei94prl1749} or Maxwell's equations~\cite{sim99jcp816} have been presented. In the former, the automaton is constructed from a moving average paradigm, while the latter is a modified version of the Lattice Gas Automaton~\cite{rot94rmp1417}.

\subsection{Cellular computing for solving PDE's}

In most cases, the predictions of calculus
based, continuous models and those of discrete, automata based ones,
are seldom quantitatively identical, though qualitative similarity is
often obtained. This is mostly explained by the fact that the two
drastically different approaches are applied to their own class of
problems.

Some attempts have recently been made to set up the solution of a PDE by using a regression method~\cite{zhou03}. The idea there is to measure an error at each discrete point of the system, and to drive an optimization process in order to find the continuous function that minimizes  this error, this function being taken in a parametrized set of continuous functions defined by a multi-layer perceptron.  This error is null if the function that is found meets the EDP requirements. Such regression processes, based on classical empirical risk minimization, are known to be sensitive to over fitting~\cite{vapnik00}.


Other attempts at a quantitative link have however been made by
showing connections between an automaton and a particular differential
problem~\cite{tok96prl3247} or by designing methods for describing
automata by differential
equations~\cite{doe05mcm977,kun04jpsj2033,omo84pd128} allowing in the
way to \emph{assess} the performance of two different
\emph{implementations} of the same problem, which are in fact basically
two different automata for the description of the same physics.

The interest of solving PDEs with cellular automata is of course not limited to physics, since PDEs are also intensively used in image processing~\cite{aubert06}. Some cellular-based solutions have also been proposed in that field~\cite{Rekeczky02}. This stresses the need for generic tools for simulating PDEs in many areas. In~\cite{Rekeczky02}, an attempt has been made to provide ready-to-use programming templates for the design of cellular algorithms, and previously mentioned software~\cite{loncar00} help to rationalize this design for PDEs.

In this paper, the problem of designing a cellular algorithm from a given partial differential problem is addressed in an attempt to bridge the present gap~\cite{bandmann02} between continuous PDE and discrete cells. To this aim, we have adapted the Least Squares Finite Elements Method \cite{Jiang} (LSFEM). In the following, section \ref{sec:scheme} describes our adaptation of the LSFE Method. Section \ref{sec:local} shows that the proposed algorithm can be made purely local and thus implemented thanks to cellular computing. Section \ref{sec:appli} provides implementation and application examples.




%
%
%
\section{Adaptation of LSFEM to cellular computing}
\label{sec:scheme}
In the following, we will reformulate the LSFE Method in a mathematical formalism which can then be  used in a cellular scheme. Subsection \ref{subsec:definitions} sets the necessary mathematical basis where one particular state of a cellular network is viewed as a function from a discrete set (the cells) into a vector space (each cell hosts a vector of reals). Subsection \ref{subsec:general} describes how the LSFEM functional is set and minimized. Finally, subsection \ref{subsec:localonly} suggests that stochastic gradient descent~\cite{Spall03} can help at making the computations local only. This will be proved in section \ref{sec:local}.

Unfortunately, the necessary mathematical formalism used in this section can seem quite abstract. To overcome this difficulty, we will provide, at each step, a simple example: a normalized mono dimensional Poisson equation, $\triangle \EVSym \left(x\right)=\frac{\partial^2\EVSym}{\partial x^2}=\ChargeSym\left(x\right)$, $\EVSym$ being the unknown electrostatic potential and $\ChargeSym$ a given repartition of charges.  The example chosen has of course a straightforward solution but it is simple enough so that each step can be detailed in the paper.

\subsection{Definitions}
\label{subsec:definitions}
The very characteristic of continuous physics is its intensive use of \emph{fields}. If we note $\FunSet AB$ the set of functions from $A$ to $B$, a field $\FieldFunctionSym \in \FunSet {\SetSpace} {\SetField}$ is a mapping of a given vector physical quantity ---belonging to  $\SetField$--- over a given physical space $\SetSpace$, for $\left(\SpaceDim,\FieldDim\right)\in\Nat^2$. For instance, our example electrostatic potential field in a 1D space is a scalar mapping over $\Reel$, as an electric field over a 3D space would be a 3D vector mapping over $\Reel^3$. Furthermore, if time were present in this example, it would be treated equally as just an additional dimension. For instance, a time-resolved 3 dimensional problem is considered as having 4 dimensions.

Therefore, a particular local differential problem $\DiffProb$ stemming from local relationships,
can be expressed in terms of a functional equation
$\PDE{\FieldFunctionSym}=0$, where the
field $\FieldFunctionSym$ is the unknown, and where $\PDESym$ represents the differential relationships derived from physical considerations, that a field $\FieldFunctionSym$ should satisfy to be the solution of $\DiffProb$. Let us note here that the functional equation $\PDE{\FieldFunctionSym}=0$ merely represents any differential equation, or system of equations, over a field $\FieldFunctionSym$ of one or more dimensions.
In our example, the field $\FieldFunctionSym$ is the association of an electrostatic potential $\Apply \EVSym x$ to each point $x$. The equation that has to be satisfied is $\forall x,\; \triangle \Apply \EVSym x - \Apply \ChargeSym x=0$. This is better expressed by the corresponding functional equation, $\triangle \EVSym-\ChargeSym=0$, where the whole mapping $\EVSym$ is the unknown.

 $\PDESym$ can thus be defined as follows, where $\LawDim\in\Nat$ can be thought of as the number of independent real equations necessary to express the local relationships which are to be satisfied at any point of $\SetSpace$ ($\LawDim=1$ in our example since only one scalar equation describes the problem):

\[
\FunDef \PDESym {\FunSet {\SetSpace} {\SetField}}
{\FunSet {\SetSpace} {\SetLaw}} \FieldFunctionSym {\PDE \FieldFunctionSym}
.
\]

In other words, $\ApplyFun \PDESym \FieldFunctionSym$ is a mapping of a vector of $\LawDim$ real values over the physical space $\SetSpace$. For any point $x\in\SetSpace$ in space, the $i^{\mathrm{th}}$ component of $\Apply {\ApplyFun \PDESym \FieldFunctionSym} x \in \SetLaw$, which would be zero if $\FieldFunctionSym$ was the solution of $\DiffProb$, actually corresponds to the local amount of violation of the $i^{\mathrm{th}}$ real local equation used to describe the problem at $x$: in our example, $\triangle\EVSym\left(x\right)-\ChargeSym\left(x\right)$, if not null, is the violation of Poisson equation at $x$. This is the heart of LSFEM: all violations, or errors, on all points can be summed up to a global error, which can itself be minimized. This is developed in the next paragraphs.

 Using a functional equation instead of considering $\FieldFunctionSym$ as a given numerical instantiation as is usually done, allows to point out  that the differential problem expressed by the mapping $\PDESym$ depends intrinsically on the unknown field $\FieldFunctionSym$, whatever its actual instantiation, or value, is.
The functional formalism allows to handle the dependency itself, \emph{i.e.} the way all violations $\ApplyFun \PDESym \FieldFunctionSym$ over the physical space $\SetSpace$  depend on the whole field $\FieldFunctionSym$.

With these notations, finding the solution to the differential problem $\PDESym$ means finding $\Opt\FieldFunctionSym$ for which $\ApplyFun\PDESym {\Opt\FieldFunctionSym}=0$. This can be done by conventional approaches such as the multidimensional Newton minimization method or well known gradient descent such as conjugate gradient. All such methods could be used in the framework of our paper, each having its own advantages and disadvantages. For the sake of clarity and illustration purposes, we have chosen to develop our paper on the Newton Minimization Method but all concepts and demonstrations can be generalized to the other minimization methods.

 We will express this method using functional derivatives of the mapping $\PDESym$ with respect to $\FieldFunctionSym$ to set the basis for understanding how we can make it local only.
Let us note here however that the functional derivatives are not as mathematically exotic as they may seem: they simply correspond to the derivative of one side of a differential equation with respect to the unknown field itself. In our example, this means deriving $\triangle \EVSym-\ChargeSym$ with respect to $\EVSym$.

To make this approach computationally tractable, we need to discretize the problem. This is performed by discretizing $\PDESym$ on a finite mesh $\Maillage \subset \SetSpace$, the discretized problem being then expressed as $\DPDE{\DFieldFunctionSym}=0$, where $\DFieldFunctionSym$ is the unknown and $\DPDESym$ is defined as follows:
\begin{equation}
\FunDef \DPDESym { \FunSet{\Maillage}{\SetField}}
 {\FunSet{\Maillage}{\SetLaw}} \DFieldFunctionSym {\DPDE \DFieldFunctionSym}
\label{eq:DDiffProb}
.
\end{equation}

We will not address, in this paper, the difficult question of the optimum mesh $\Maillage$ which allows the discrete solution to be the closest to the continuous one. We will thus assume that $\Maillage$ is correctly chosen with respect to the differential problem itself so that conventional methods would give satisfactory results.


In contrast, the question of the treatment of boundary conditions, whether they be of the Dirichlet or Neumann type is of primary importance. A Dirichlet condition expressed as some $\Maille \in \Maillage$ is to expressed as $ \forall \DFieldFunctionSym,\, \Apply {\DPDE \DFieldFunctionSym} \Maille = 0$: in other words, the restriction of $\DPDESym$ to $\Set{\Maille}$ is null. Equally, specifying that Neumann conditions are to be satisfied on a subset of $\Maillage$ is equivalent to giving a specific definition of $\DPDESym$ over this subset. This idea can be further generalized by considering several subsets of $\Maillage$ over which the general expression of $\DPDESym$ changes. This would allow to take into account subdomains of $\Maillage$, each of which having its own differential problem. However and whatever the precise differential problem and thus the actual definition of $\DPDESym$, this shows that boundary conditions are not to be \emph{added} to the discretised differential problem $\DPDESym$ but are inherently \emph{part of it}, as it should mathematically be.

To clarify this, let us go back to our example: the solving of the 1D Poisson equation on a mono dimensional mesh $\Maillage$ of $\ExampleN$ regularly spaced points $x_1,\cdots, x_\ExampleN$. In the following, the value $\Apply \EVSym {x_i}$ associated by the mapping $\EVSym$ at the point $x_i$ will be shortened to $\EVati$. The same stands for the charge $\Apply \ChargeSym {x_i}$ at the point $x_i$, that will be written as $\ChargeAt i$. The discretized problem can then be found by finite difference as the following, provided $\EVat 1$ and $\EVat\ExampleN$ are defined as Dirichlet boundary conditions and $\ExampleStep$ is the sampling step : $$\forall i\in\Nat, 1<i<\ExampleN, \frac{1}{\ExampleStep^2}\left(\EVatim 1-2\EVati+\EVatip 1\right)-\ChargeAt i=0.$$
Once again, let us us stress here that the whole expression, including all space points is seen as depending on a {\em single} functional parameter $\EVSym$, which is a function over the discrete set $\Set{x_1,\cdots,x_\ExampleN}$. This function $\EVSym$ is what is actually generally formalized above as $\DFieldFunctionSym$.

\subsection{General method}
\label{subsec:general}

Getting back to the general case and as was already discussed, solving the problem means finding $\Opt\FieldFunctionSym$ for which $\ApplyFun\PDESym {\Opt\FieldFunctionSym}=0$, which
means finding a field $\DFieldFunctionSym$ for which
$\DPDE\DFieldFunctionSym$ is as close to the $0$ mapping as possible given a distance on the functional space
${\FunSet{\Maillage}{\SetLaw}}$. This, in turn, is equivalent to
zeroing all $\LawDim$ relations $\Apply { {\DPDE \DFieldFunctionSym}} {\Maille}$ for
\emph{all} $\Maille\in\Maillage$.
Finally, this can be equivalently done by similarly minimizing a the functional expression  $\Err\DFieldFunctionSym$ as is done in the LSFE Method \cite{Jiang}:
\begin{equation}
\Err\DFieldFunctionSym=\sum_{\Maille \in \Maillage} {\norm{\Apply {
{\DPDE \DFieldFunctionSym}} {\Maille}}}
\label{eq:error}
\end{equation}
 where $\norm\quad$ is any given norm on $\SetLaw$. The usual LSFEM continuous integral is here replaced by a discrete sum because we have already discretized the differential problem so as to formalize the use of cellular computing. Strictly speaking, we depart here from the Least Squares Finite Element Method and should rather call our method a Least Squares Finite Difference Method.

In our example, if the norm is chosen as the simple square, equation (\ref{eq:error}) translates to
$$\Err\EVSym=\sum_{i=2}^{N-1}\left( \frac{1}{\ExampleStep^2}\left(\EVatim 1-2\EVati+\EVatip 1\right)-\ChargeAt i\right)^2.$$

As mentioned previously, we have to set $\DPDESym$ so that it {\em includes} the satisfaction of the differential equations at boundary conditions. This has been done here easily for the Dirichlet type by just a priori removing boundary  terms $1$ and $\ExampleN$ from the sum, because their values are known from the Dirichlet conditions and thus no error can be committed on them.

Now that the error functional $\Err\DFieldFunctionSym$ is defined, the LSFE Method prescribes that it be minimized so as to find the value $\Opt\DStateSym$ which produces the best solution to the initial problem. This can be done by numerous numerical methods such as the steepest or conjugate gradient (see for instance \cite{Bishop}). As we chose not to restrain our study to a specific differential problem, we have no particular reason to chose one particular minimization method. Thus, for illustration and demonstration purposes, we have chosen the standard Newton minimization method applied to multidimensional problems. The following considerations are valid whatever the method chosen.
Let us note here however that this minimization process does not ensure the zeroing of $\Err\DFieldFunctionSym$, which is to be verified \emph{a posteriori} by evaluating $\Err{\Opt\DStateSym}$.

The computation of $\Err\DFieldFunctionSym$ produces a scalar from a given state $\DFieldFunctionSym$ of the discretized problem variables. This scalar can be viewed as an {\em evaluation} of this state. For further purpose, let us define more generally an evaluation as a function $\EvaluationSym \in\FunSet{\FunSet{\Maillage}{\SetField}}{\Reel}$. $\ErrSym$ is precisely an {\em evaluation} that is suited for quantifying the quality of a particular instantiation of $\DFieldFunctionSym$ as a solution to the discretized differential problem  $\DPDESym$ .

To undertake this optimization task, we previously need to define a canonical basis of  the functional space ${\FunSet{\Maillage}{\SetField}}$ with respect to which the \emph{gradient} and \emph{Hessian} will be taken. This basis is the set of the Cellular Network states in which each state is totally null except one given component of one given cell, which is set to 1. The number of basis elements is thus equal to the number of cells multiplied by the number of reals in each cell. This is mathematically defined as the following:
if $\kroneckerSym$ is the Kronecker symbol and $\Serie\CanonReelSym i$ is the  canonical basis  of $\SetField$, let us define $\CanonBase$, the canonical basis of ${\FunSet{\Maillage}{\SetField}}$ as the set of functions $\CanonFunc\Maille i$, for all $\Maille\in\Maillage$ and all $1 \le \left(i\in\Nat\right)\le\FieldDim$:
\begin{equation}
\FunDef{\CanonFunc{\Maille}{i}}{\Maillage}{\SetField}{\mMaille}{\kronecker\Maille\mMaille\SerieElem\CanonReelSym i}
\label{eq:base}
\end{equation}

The partial derivative of an evaluation $\EvaluationSym$ at point
$\DFieldFunctionSym$ according to basis vector $\CanonFunc\Maille i$ is by
definition $\lim_{h\in\Reel\rightarrow 0}\left(\Apply \EvaluationSym
{\DFieldFunctionSym+h\CanonFunc\Maille i} - \Apply \EvaluationSym
\DFieldFunctionSym\right)/h$. This value is, by definition of the gradient, the actual $(\Maille,i)$ component of $\Grad {\EvaluationSym} {\CanonBase}  \DStateSym$.



In our example, the basis vector $\CanonFunc\Maille i$ is reduced to $\CanonFunc\Maille 1$ since $\LawDim=1$. As $\omega$ is a given $x_i$, this basis vector is the mapping with 0 potential everywhere, except at $x_i$ where the value $\EVati$ equals $1$. Let us write this $\CanonFunc{x_i} 1$ as $\ElecBase i$ for our example.

Using these definitions of derivation and getting back to the general case, the Newton method consists in building a series $\Timet \DStateSym$ defined as follows, the limit of which should be the sought solution $\Opt\DFieldFunctionSym$ to $\DDiffProb$, the field which is the solution of our initial differential problem:

\begin{equation}
\begin{array}{rcl}
\Timetpu \DFieldFunctionSym & = & \Timet \DFieldFunctionSym -
\FunMAJ {\ErrSym} {\CanonBase} {\Timet \DStateSym} \\

\FunMAJ {\ErrSym} {\CanonBase} {\Timet \DStateSym}
& = &
{\HessInv {\ErrSym} {\CanonBase} {\Timet \DStateSym}}.{\Grad {\ErrSym} {\CanonBase} {\Timet \DStateSym}} \\ ~ \\
\multicolumn{3}{l}{\Ou \; \HessInvSym \mbox{ is the inverse of the Hessian matrix.}}
\end{array}
\label{eq:maj}
\end{equation}

The above expression requires some derivability conditions on $\ErrSym$, and thus on both $\DPDESym$ and the chosen norm on $\SetField$. The former is assumed, since it stems from the problem $\DiffProb$ itself: the differential problem is here assumed to be derivable with respect to the unknown field. The latter is ensured by the appropriate choice of the used norm. As another precaution to be taken on that choice, the used norm must ensure that
no component of the gradient -- and thus of the Hessian inverse -- neither supersedes the others nor is superseded by them, for this is known to create stability problems in the iteration defined by (\ref{eq:maj}). The conventional $\norm{\quad}_2$ norm, or its square, is for instance a good choice, provided $\DiffProb$ is conveniently normalized, \emph{i.e.} that the unknown of the initial differential problem is a normalized quantity which has an order of magnitude around 1.

Equation (\ref{eq:maj}) can be applied to our example by simply replacing ${\Timet \DStateSym}$ by $\Timet \EVSym$ and $\CanonBase$ by the set of all $\Serie\ElecBaseSym i$. This yields a complex expression for $\FunMAJ {\ErrSym} {\Serie \ElecBaseSym i} {\SerieElem \EVSym \StepIdx}$, too complicated too show here, that involves all $\EVati, 1\le i\le\ExampleN$.

\subsection{Local only computations}
\label{subsec:localonly}

The effective computation of
such a series as defined by (\ref{eq:maj}) implies to compute, for each step $\StepIdx$, the gradient and inverse Hessian with respect to $\CanonBase$, which implies getting access to the whole $\Maillage$. This is in contradiction with our initial goal which was to design a  computational method which can be implemented in a cellular way. Indeed, this requires that the method be \emph{local-only}. This means that the evaluation of a particular cell of the mesh requires only the knowledge of the values in a few neighboring cells instead of the whole $\Maillage$.

To overcome this limitation, we present in the following a method inspired from the stochastic
gradient descent method~\cite{Spall03}, the locality of which will be
established in the next section.

 The stochastic gradient method consists in updating  $\DStateSym$ by considering only a few of its components at a time. We choose to consider a single $\Maille$ in $\Maillage$ at each step, thus modifying only $\Component \DStateSym \Maille$, the $\Maille$-related components of $\DStateSym$, \emph{i.e.} the values of the field at a given point in the mesh. Therefore, the gradient and Hessian appearing in (\ref{eq:maj}) are taken not with respect to the whole $\CanonBase$ but rather with a subset $\LocCanonBase \Maille$ of it, restricted to $\Maille$, defined as the set $\Set{\CanonFunc\mMaille i: \mMaille = \Maille \Et 1\leq i \leq \FieldDim}$.

The system of interdependent equations resulting from the problem discretization is thus derived with respect to the field values at a given point at a time only.
One such step is therefore defined as follows:

\begin{equation}
\Timetpu{\Component \DStateSym \Maille} = \Timet{\Component \DStateSym \Maille} - \FunMAJ {\ErrSym} {\LocCanonBase\Maille} {\Timet \DStateSym}
\label{eq:locmaj}
,
\end{equation}
which, in the frame of our example, translates to:
$$
\Timetpu{\EVati} = \Timet{\EVati} - \FunMAJ {\ErrSym} {\ElecBase i} {\Timet{\EVSym}}
.
$$

\begin{table*}
 \center
$$
 \begin{array}{|l|c|}
\hline
   i=1\quad \mathrm{or}\quad i=\ExampleN& \Timetpu {\EVati} = \Timet {\EVati}\\
\hline
  i=2 & \Timetpu{\EVati} =\frac{1}{5}\left(2\Timet{\EVatim 1}+4\Timet{\EVatip 1}-\Timet{\EVatip 2}+\ExampleStep^2\times\left(\ChargeAt{i+1}-2\ChargeAt{i}\right)\right)\\
\hline
  i=\ExampleN-1 & \Timetpu{\EVati}=\frac{1}{5}\left(2\Timet{\EVatip 1}+4\Timet{\EVatim 1}-\Timet{\EVatim 2}+\ExampleStep^2\times\left(\ChargeAt{i-1}-2\ChargeAt{i}\right)\right)\\
\hline
  3\le i\le \ExampleN-2 & \Timetpu{\EVati}=\frac{1}{6}\left(-\Timet{\EVatim 2}+4\Timet{\EVatim 1}+4\Timet{\EVatip 1}-\Timet{\EVatip 2}+{\ExampleStep^2}\times\left(\ChargeAt{i-1}-2\ChargeAt{i}+\ChargeAt{i+1}\right)\right)\\
\hline
 \end{array}
$$
\caption{Update rules for the mono dimensional automaton which solves the mono dimensional Poisson equation, as computed from (\ref{eq:locmaj}).}
\label{tab:MAJ}
\end{table*}

The above relationship describes a series for a given point $\Maille$ of $\Maillage$. For the series (\ref{eq:maj}) to be completely approximated by the stochastic method, the relationship (\ref{eq:locmaj}) is to be iterated over $\Maillage$ with a random choice of $\Maille \in \Maillage$ at each step: for the derivative to be complete, it is here taken successively with respect to the field values at each point in the mesh.

Thus, provided $\AbstractFunMAJ {\ErrSym} {\LocCanonBase\Maille}$  is somehow local, an issue that will be addressed in the next section, the above considerations allow to consider (\ref{eq:locmaj}), at $\Maille$, as the definition of a \emph{continuous} automaton, which is an extension of classical cellular automata for which the cell states are allowed to take their values in $\SetField$. This automaton can be implemented for any given differential problem $\DiffProb$ by evaluating $\AbstractFunMAJ {\ErrSym} {\LocCanonBase\Maille} $ for this particular problem.

Evaluating $\AbstractFunMAJ {\ErrSym} {\LocCanonBase\Maille}$ can be done, as equation (\ref{eq:maj}) suggests, by taking the proper gradient and Hessian of the discretized problem at each point in the mesh. Applied to our example, this method allows to calculate, for each point in the mono dimensional mesh, the update rule to be applied to that point. 4 different rules are found, which are given in table \ref{tab:MAJ}.

As can also be inferred from table \ref{tab:MAJ}, the  automaton described by (\ref{eq:locmaj}) for each $\Maille$  departs from the strict definition of a cellular automaton by the fact that the update rule for all cells $\Maille$ are only the same for a vast majority of them, but not strictly all. Indeed, because of the existing  boundary conditions, the $\HessSym$ and $\GradSym$ operators will not give the same result for all points, since the boundary conditions are considered as constants. Hence, a Dirichlet boundary is described in the automaton by a constant cell, the value of which is given by the automaton initial state.

At this point of the paper, we have defined a cellular algorithm that can be automatically generated from any given differential problem, thanks to automated formal derivative computing, by evaluating the stochastic gradient descent  ({\ref{eq:locmaj}}) at each point of the mesh. We have done so by assuming that the update rules thus computed is local. To formally establish that the generated algorithm is indeed cellular, we now need a proof of this assumption. It is the subject of the next section.


\section{Localization of each cell neighborhood}
\label{sec:local}
The demonstration of the locality of the variant of the LSFE Method we have presented is a key point in this paper, as cellular computing which can be implemented on parallel architecture is our essential goal. It is done in a two step procedure detailed in the next two subsections. The first step is the definition of the neighborhood of a given cell $\Maille$ in the mesh:  it is the set of the cells whose values are needed in the computation of the update of $\Maille$.

The second step consists in evaluating the size of this neighborhood by determining which cells are elements of it. This demonstration is done by considering the particular Newton method for minimization but it would be valid for any other method as only the properties of the derivatives are used.

\subsection{Neighborhood definition}
The definition of the \emph{neighborhood} $\Apply \FunVoisSym \EvaluationSym$ of a given {\em evaluation} $\EvaluationSym$ (see section~\ref{subsec:general} for a definition) is thus to be understood as being the set of all the points $\Maille$ needed in the computation of $\EvaluationSym$.

\begin{equation}
\FunDef \FunVoisSym {\FunSet{\FunSet{\Maillage}{\SetField}}{\Reel}} {\Parties \Maillage} {\EvaluationSym} {
\Set{\Maille \in \Maillage 
\Tq  \Exists \DStateSym \Tq \Grad {\EvaluationSym} {\LocCanonBase\Maille} {\DStateSym} \neq 0}}
\label{eq:defvois}
\end{equation}

\begin{table}
 \center
$$
 \begin{array}{|l|c|}
\hline
  i=1\quad\mathrm{or}\quad i=\ExampleN& \{x_{i}\}\\
\hline
  i=2 & \{x_{i-1},x_{i+1},x_{i+2}\}\\
\hline
  i=\ExampleN-1 & \{x_{i-2},x_{i-1},x_{i+1}\}\\
\hline
  3\le i\le \ExampleN-2 & \{x_{i-2},x_{i-1},x_{i+1},x_{i+2}\}\\
\hline
 \end{array}
$$
\caption{Neighborhoods $\Vois {\norm {\AbstractFunMAJ {\ErrSym} {\Serie \ElecBaseSym i}}}$ for all cells of an automaton which solves the mono dimensional Poisson Equation}
\label{tab:vois}
\end{table}

The algorithm described in the previous section is thus practically usable if the calculations needed to evaluate each cell are indeed \emph{local}, \emph{i.e.} expression (\ref{eq:locmaj}) can be evaluated without requiring access to $\Maillage$ as a whole. This can be formally stated as $\Vois {\norm {\AbstractFunMAJ {\ErrSym} {\LocCanonBase\Maille}}}\ne\Maillage$.
This can happen only if some kind of locality condition on $\DiffProb$ is assumed, \emph{i.e.} if the initial differential problem is expressed in a local manner, as it is usually the case.

In the frame of our example, the values of $\Vois {\norm {\AbstractFunMAJ {\ErrSym} {\ElecBase i}}}$ for all points in the mesh are given in table \ref{tab:vois}. For instance, the last row of table
table~\ref{tab:MAJ} show that the value of $\EVSym$ at points $x_{i-2}$, $x_{i-1}$, $x_{i+1}$ and $x_{i+2}$ are required to evaluate $x_i$, thus leading to the neighborhood $\{x_{i-2},x_{i-1},x_{i+1},x_{i+2}\}$ shown on the last row of table \ref{tab:vois}.

\subsection{Neighborhood size}

To show that the automaton is indeed local in the general case, let us first consider the specific case of the evaluation $\norm {\Apply \DPDESym \Maille}$, that is the error measurement at point $\Maille$. The global error evaluation $\ErrSym$ is a summation of such terms (see~(\ref{eq:error})).

For further use, we now need to define an enhancement of the neighborhood concept we have called the \emph{dependency} $\nVois\Maille\DPDESym$ of a given $\Maille\in\Maillage$ involved in a problem $\DPDESym$. It is the set of point $\mMaille$ for which $\Maille$ belongs to the neighborhood of $\mMaille$:
$$
\nVois\Maille\DPDESym = \Set{\mMaille \Tq \Maille \in \Vois {\norm{\Apply { {\DPDESym}} {\mMaille}}}}.
$$
Given the definition (\ref{eq:maj}) of $\AbstractFunMAJ {\ErrSym} {\LocCanonBase\Maille}$, the gradient can be linearly distributed over the additive components of $\ErrSym$ as in~(\ref{eq:simplify2}).
The summation term appearing in~(\ref{eq:simplify1}) has been restricted to those $\mMaille$ in $\Maillage$ for which the gradient does not vanish, {\em i.e.} those $\mMaille \in \nVois\Maille\DPDESym$. The summations product in~(\ref{eq:simplify2}) is obtained by similarly distributing the Hessian.

\begin{eqnarray}
\AbstractFunMAJ {\ErrSym} {\LocCanonBase\Maille}
  = 
\sum_{\mMaille\in\nVois\Maille\DPDESym}{{\AbstractHessInv {{\norm{\ErrSym}}} {\LocCanonBase\Maille}}\AbstractGrad {{\norm{\Apply { {\DPDESym}} {\mMaille}}}} {\LocCanonBase\Maille}} \label{eq:simplify1}\\
  = {\AbstractHessInv {{\norm{\ErrSym}}} {\LocCanonBase\Maille}}\sum_{\mMaille\in\nVois\Maille\DPDESym}{\AbstractGrad {{\norm{\Apply { {\DPDESym}} {\mMaille}}}} {\LocCanonBase\Maille}} \nonumber\\
  = 
\Inv{\sum_{\mMaille\in\nVois\Maille\DPDESym}{{\AbstractHess {{\norm{\Apply { {\DPDESym}} {\mMaille}}}} {\LocCanonBase\Maille}}}}\sum_{\mMaille\in\nVois\Maille\DPDESym}\AbstractGrad {{\norm{\Apply { {\DPDESym}} {\mMaille}}}} {\LocCanonBase\Maille}\label{eq:simplify2}
\end{eqnarray}

The neighborhood of a product being included in the union of its operands neighborhoods, from (\ref{eq:defvois}), the neighborhood of $\norm{\AbstractFunMAJ {\ErrSym} {\LocCanonBase\Maille}}$, according to (\ref{eq:simplify2}), can be limited to
\begin{eqnarray}
\Vois {\norm{\AbstractFunMAJ {\ErrSym} {\LocCanonBase\Maille}}}
\subset
\VoisLight{\norm{\Inv{\sum_{\mMaille\in\nVois\Maille\DPDESym}{\AbstractHess {{\norm{\Apply { {\DPDESym}} {\mMaille}}}} {\LocCanonBase\Maille}}}}} \nonumber \\
\bigcup\VoisLight{\norm{\sum_{\mMaille\in\nVois\Maille\DPDESym}{\AbstractGrad {{\norm{\Apply { {\DPDESym}} {\mMaille}}}} {\LocCanonBase\Maille}}}} 
\label{BigUnion}
\end{eqnarray}

From definition (\ref{eq:defvois}), it can be shown that the neighborhood of a derivative, or a gradient, is included in the neighborhood of its operand. The same holds for the neighborhood of a Hessian since any line or column of the Hessian is a derivative of the gradient. Therefore, the right-hand term of the above union is included in $\bigcup_{\mMaille\in\nVois\Maille\DPDESym}\Vois{{\norm{\Apply { {\DPDESym}} {\mMaille}}}}$.

Furthermore, the neighborhood of a matrix norm $\norm{M}$ is obviously included in the union of the neighborhoods of all its components. The same holds for the inverse matrix $\norm{\Inv{M}}$ since each of its components can be obtained by a combination of the components of $M$. We can therefore conclude that the left-hand term of the union in (\ref{BigUnion}) is also a subset of $\bigcup_{\mMaille\in\nVois\Maille\DPDESym}\Vois{{\norm{\Apply { {\DPDESym}} {\mMaille}}}}$.

Therefore, provided we can assume that $\Vois{{\norm{\Apply { {\DPDESym}} {\Maille}}}}$ is small enough for all $\Maille\in\Maillage$ ---which is ensured if the differential problem $\DiffProb$ is defined locally---, the calculations to be undertaken to evaluate
$\AbstractFunMAJ {\ErrSym} {\LocCanonBase\Maille}$ for each cell $\Maille$ are local to some extended neighborhood of that cell:
\[
\Vois {\norm{\AbstractFunMAJ {\ErrSym} {\LocCanonBase\Maille}}}
\subset
\bigcup_{\mMaille\in\nVois\Maille\DPDESym}\Vois{{\norm{\Apply { {\DPDESym}} {\mMaille}}}}
\]

From the definition of the neighborhood, the above inclusion means that the calculations involved in computing the update rule $\AbstractFunMAJ {\ErrSym} {\LocCanonBase\Maille}$ at a given point in the mesh only involve the field values of the dependent points $\mMaille$ in the sense of $\nVoisSym$, the actual number and repartition of those points being dependent on the differential problem itself.
(In the frame of our example, the automaton obtained by our formal resolution process is given in table \ref{tab:MAJ}.)


We have now proven that the stochastic gradient descent applied to the Newton minimization in LSFEM can be implemented through cellular computing, provided that the initial differential problem is itself local as it is generally the case.  This is particularly interesting if a programming cellular environment, analogous to those described in \cite{can94pc803,spe01book51}, is available not only on shared memory multi-processor computers but also on distributed memory architectures such as clusters \cite{GUSTEDT:2006:INRIA-00103772:1,GUSTEDT:2007:HAL-00280094:1,FRESSENGEAS:2007:INRIA-00139660:1}.

\section{Application examples}
\label{sec:appli}

This next section is thus devoted to the presentation of application examples on two classical Dirichlet boundary value problems. We will however not present any performance analysis in terms of computing time until convergence since this highly depends on the actual parallel implementation of the cellular algorithms \cite{GUSTEDT:2006:INRIA-00103772:1,GUSTEDT:2007:HAL-00280094:1}. This work is currently in progress and an analysis of the obtained computing time has already been done \cite{FRESSENGEAS:2007:INRIA-00139660:1}.

As hinted before, we have implemented the continuous automaton described in the previous sections with the help of an off-the-shelf formal computing software\footnote{We have used \texttt{SAGE} Mathematical Software, Version 4, \url{http://www.sagemath.org}}, essentially used to formally evaluate the update rule from the differential problem through equation (\ref{eq:locmaj}), and a cellular automata environment analogous to those reported in~\cite{can94pc803,spe01book51}. We have thus automated the computation from the specification of the discrete differential problem $\DDiffProb$ to the design of the adequate continuous automaton solving the differential problem through LSFEM\footnote{The corresponding piece of software is available under GPL license on the authors web sites.}.

Let us now illustrate this process with two examples. The first one is the generalization to 3 dimensions of the example used in the previous sections. The mono dimensional example was trivial, as it possessed a straightforward solution. The 3D one is a little trickier as it is a 3D boundary value problem. The second example is the application of strictly the same piece of software to the non-paraxial beam propagation equation, which is not so easy to solve numerically.

\subsection{Poisson equation}

The first example  is thus the solving of a normalized Poisson Equation for $\EVSym$ in the three dimensions of space: $\triangle \EVSym\left(x,y,z\right)=\ChargeSym\left(x,y,z\right)$ for any given $\ChargeSym$, the Dirichlet boundary conditions being set on the sides of the computing cube window. The corresponding discrete problem is straightforward and is obtained through finite difference centered second derivatives on each dimensions of space, for the same space step $\SpaceStep$.

The automaton obtained through the evaluation of (\ref{eq:locmaj}) on each point of the mesh has $28$ different update rules. The update rule obtained for $\Maille$ such as the boundaries conditions are not in $\Vois {\norm{\AbstractFunMAJ {\ErrSym} {\LocCanonBase\Maille}}}$ concerns the vast majority of the mesh nodes $\Maille$ and is shown below. It is a centro-symmetric three dimensional convolution kernel involving $\EVSym$ and $\ChargeSym$. Only middle and lower parts of these kernels are shown, the upper part being obtained by symmetry.

\[
	\EVSym\leftarrow
	\frac{1}{42}
	\left[
	\EVSym\left(
\begin{minipage}{2.5cm}\includegraphics[width=2.5cm]{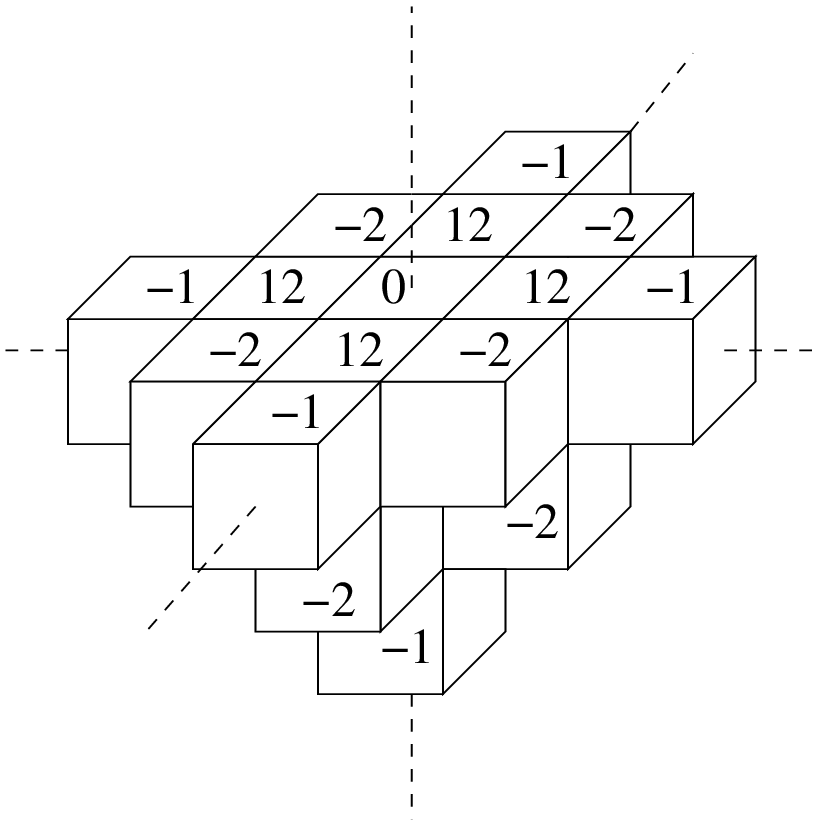}\end{minipage}
\right)
	+\SpaceStep^2.\ChargeSym\left(
\begin{minipage}{1.5cm}\includegraphics[width=1.5cm]{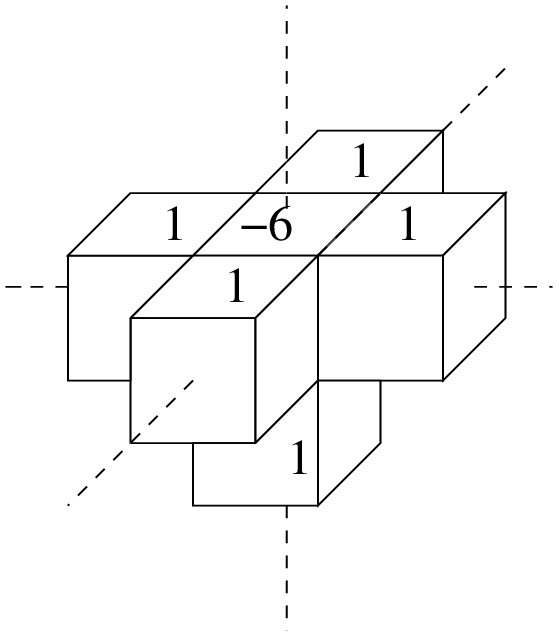}\end{minipage}
\right)
	\right]
.
\]

The 27 other update rules account for the boundary conditions.
When launched, the system converges to a fixed point, corresponding to the result that, in that case, can also be obtained with other methods.



As stated above, the result has to be checked valid \emph{a posteriori} by evaluating the remaining error as defined by (\ref{eq:error}). For a better assessment of the performance, we will provide two values of the remaining error. the first one is the \emph{mean error}, which is simply $\Err\DFieldFunctionSym$ when $\DFieldFunctionSym$ takes the value of the solution found, all divided by the number of points, to get the mean error per mesh point. The other one, the \emph{maximum error}, is defined as $\ErrMax\DFieldFunctionSym=\max_{\Maille \in \Maillage} {\norm{\Apply {{\DPDE \DFieldFunctionSym}} {\Maille}}}$ and yields the maximum error per point. Both will be normalized to the maximum component of $\DFieldFunctionSym$.

For a $20\times20\times20$ mesh and for a value of $\ChargeSym$ varying from 1 to 0 from one side of the cube to the other, the \emph{a posteriori} computed mean and maximum errors are  $7\times10^{-3}$  and $4\times10^{-1}$ respectively for $\SpaceStep=1$ and decrease with it. The maximum error, that can seem large, is due to strong gradients in the solution close to the boundary conditions and the very crude mesh used. The strong gradients are caused by the non realistic values taken for $\ChargeSym$. However, the mean error shows that the solution found, aside from a few points, is still acceptable, despite the sparse mesh.

\setlength{\columnwidth}{\linewidth}

\subsection{Non paraxial laser beam propagation}
\begin{figure*}
	\hfill \includegraphics[width=\columnwidth]{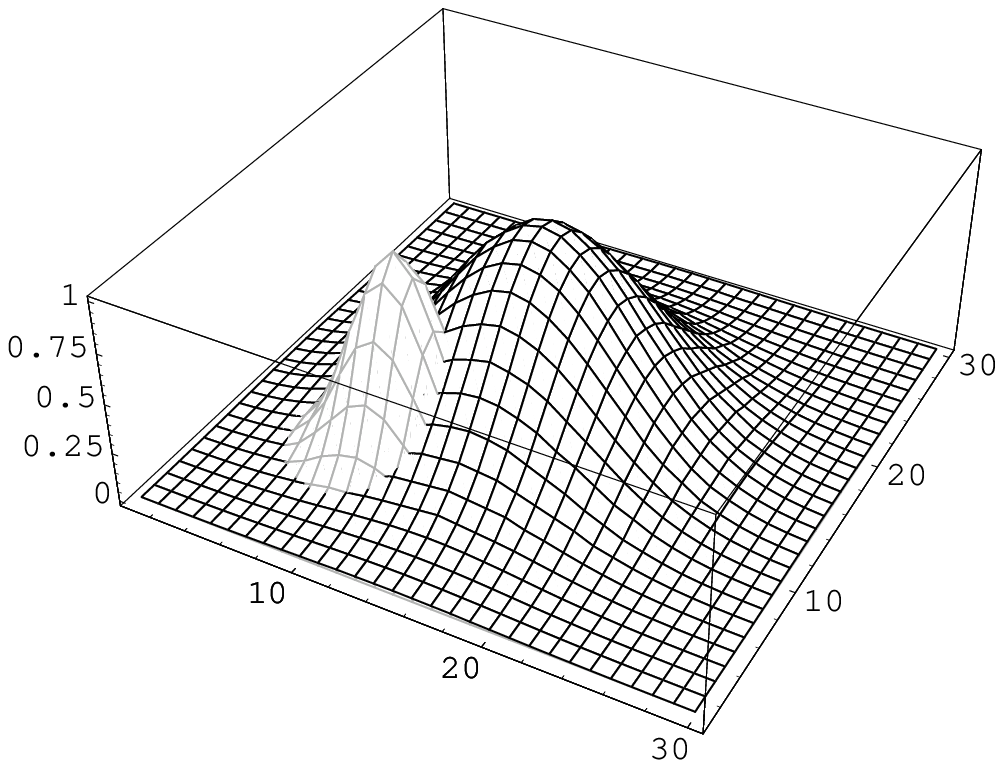} \hfill \includegraphics[width=\columnwidth]{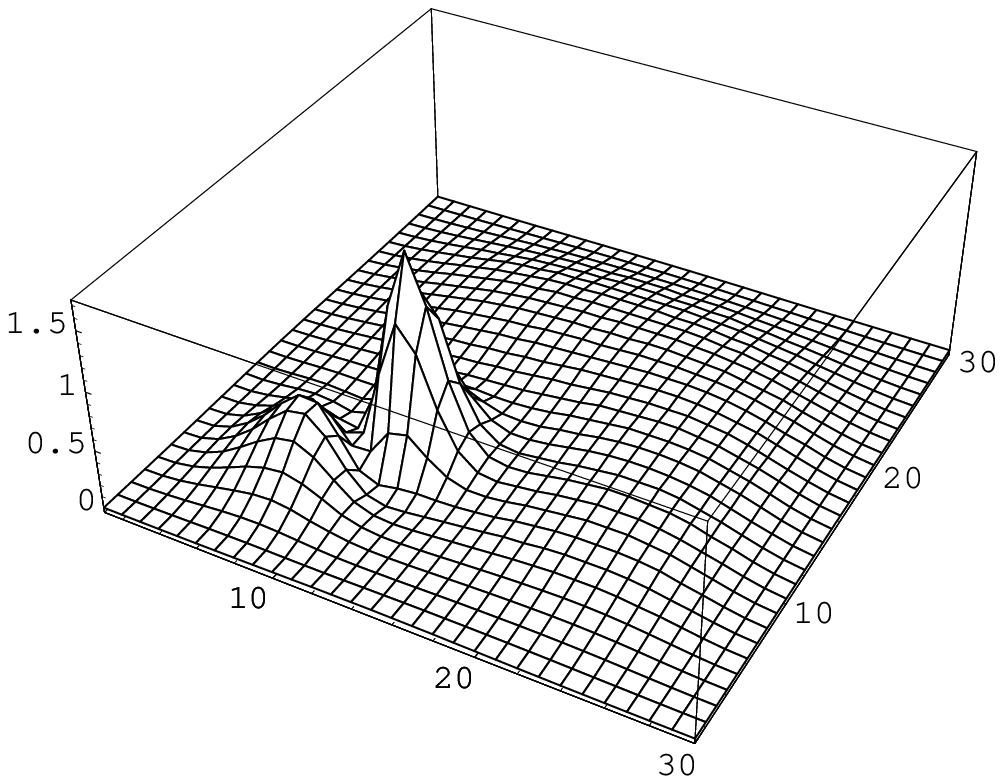} \hfill
	\caption{Left : laser beam Gaussian profile to be coupled in the waveguide (black) and waveguide (gray) (A.U.). Right : beam profile after a 3mm propagation. The window size is 30$\mu$m and the beam wavelength is 250$\mu$m. The highest peak on the right evidences the light which is coupled into the waveguide.\label{fig:wave} }
\end{figure*}

The second example is a well know difficulty in the domain of electromagnetic propagation: the removal of the paraxial approximation. Indeed, we now aim to compute the coupling of a Gaussian laser beam of width $\waist$ into a Gaussian shaped waveguide of width $\frac{ \waist}{2}$ and modulation depth $10^{-4}$. The centers of both beam and waveguide are set to a distance of $\frac{ \waist}{2}$, both being aligned in the same direction. In the computation process, we will \emph{not} make the standard simplifying paraxial approximation, which makes our problem difficult to solve by conventional methods.

The non-paraxial propagation equation to be solved is thus the following, where $\wave$ is the wave electric field to be found, $z$ is the propagation direction, $k$ is the wave vector, $n$ and $\delta n$ are the given refraction index and a small variation of it:
\begin{equation}
\frac{\partial\wave}{\partial z}-\frac{i}{2 k}\triangle\wave=\frac{i k}{n}\delta n\wave.
\label{eq:wave}	
\end{equation}

The problem is solved by deriving two real equations from (\ref{eq:wave}), discretizing them with finite difference centered derivatives except along $z$ where left-handed derivatives are needed because of the impossibility to give a boundary condition on one side of the propagation axis. 

The adequate continuous automaton (it also has 28 update rules but is too complicated to show here) is then computed from the discretized problem.
When launched on a $30\times30\times30$ network, it stabilizes to a fixed point shown on figure \ref{fig:wave}, where the light is found to be coupled into the waveguide. The \emph{a posteriori} remaining mean and maximum errors are now computed to be $2\times10^{-12}$ and $9\times10^{-12}$ respectively,
proving that the obtained solution does indeed meet the differential problem requirements.

\section{Conclusion}

We have shown how the Least Squares Finite Elements Method can be adapted for its cellular implementation. This is of particular interest as cellular algorithms can be efficiently implemented on parallel hardware \cite{FRESSENGEAS:2007:INRIA-00139660:1,GUSTEDT:2006:INRIA-00103772:1,GUSTEDT:2007:HAL-00280094:1}, paving the way for the distribution of large scale differential problems on computer networks.

A side effect of the method is the possibility to automate the design of the cellular algorithm, thanks to formal computing, from the differential problem specification down to its solution, sparing the user the need to get involved in actual numerical mathematics and computer programming, thus sparing code development time.

Thus, while the method presented here does not pretend to compete with state-of-the-art numerical methods for a given well known partial differential system, it allows people from the physics community to rapidly and efficiently run their new models on distributed parallel architectures, as was shown with three examples that raise different types of numerical difficulties.

\section{Acknowledgments}

This work was supported by the InterCell MISN program of the French State to Lorraine region 2007-2013 plan. Cellular computing software implementation was done and run on hardware funded from this grant.


\input{Escapade.bbl}
\end{document}

%% file: commands.tex
\newcommand{\MailleSym}[0]{\omega}

\newcommand{\StateSym}[0]{\xi}

\newcommand{\ErrSym}[0]{{\cal E}}
\newcommand{\ErrMaxSym}[0]{{\ErrSym_M}}
\newcommand{\GradSym}[0]{\mathrm{grad}}

\newcommand{\HessSym}[0]{{\mathrm{H}}}
\newcommand{\HessInvSym}[0]{{\bar{\mathrm{H}}}}
\newcommand{\VoisSym}[0]{{\cal V}}
\newcommand{\nVoisSym}[0]{{\cal D}}
\newcommand{\MAJSym}[0]{\mu}
\newcommand{\FieldFunctionSym}[0]{\StateSym}
\newcommand{\PDESym}[0]{\Phi}
\newcommand{\DFieldFunctionSym}[0]{\tilde{\FieldFunctionSym}}
\newcommand{\DStateSym}{\DFieldFunctionSym}
\newcommand{\DPDESym}[0]{\tilde{\PDESym}}
\newcommand{\DiffProb}[0]{\mathbb P}
\newcommand{\DDiffProb}[0]{\tilde{\mathbb P}}
\newcommand{\FunSet}[2]{{\left(#2\right)}^{#1}}
\newcommand{\EvaluationSym}[0]{\zeta}
\newcommand{\FunVoisSym}[0]{\VoisSym}

\newcommand{\Apply}[2]{#1{\left(#2\right)}}
\newcommand{\ApplyFun}[2]{#1^{\left(#2\right)}}
\newcommand{\norm}[1]{\left|{#1}\right|}

\newcommand{\DPDE}[1]{\ApplyFun \DPDESym {#1}}

\newcommand{\PDE}[1]{\ApplyFun \PDESym {#1}}
\newcommand{\Reel}{{\mathbb R}}
\newcommand{\Nat}{{\mathbb N}}
\newcommand{\Maillage}[0]{\Omega}

\newcommand{\Maille}[0]{\MailleSym}
\newcommand{\mMaille}[0]{\MailleSym^\prime}

\newcommand{\Set}[1]{\left\{ #1 \right\}}

\newcommand{\Opt}[1]{{#1}^\star}

\newcommand{\Tq}{\;:\;}

\newcommand{\Exists}{\exists}
\newcommand{\Et}{\mbox{ and }}
\newcommand{\Ou}{\mbox{ where }}

\newcommand{\Err}[1]{\ApplyFun \ErrSym {#1}}
\newcommand{\ErrMax}[1]{\ApplyFun \ErrMaxSym {#1}}

\newcommand{\Parties}[1]{{{\cal P}\left( #1 \right)}}

\newcommand{\Inv}[1]{\left( #1 \right)^{-1}}
\newcommand{\Nablaoid}[4]{\Apply {\ApplyFun {{#1}_{\left|{#3}\right.}} {#2}} {#4}}
\newcommand{\AbstractNablaoid}[3]{\ApplyFun {{#1}_{\left|{#3}\right.}} {#2}}
\newcommand{\Grad}[3]{\Nablaoid {\GradSym} {#1} {#2} {#3}} 

\newcommand{\HessInv}[3]{\Nablaoid {\HessInvSym} {#1} {#2} {#3}}
\newcommand{\FunMAJ}[3]{\Nablaoid {\MAJSym} {#1} {#2} {#3}}
\newcommand{\AbstractFunMAJ}[2]{\AbstractNablaoid {\MAJSym} {#1} {#2}}
\newcommand{\AbstractGrad}[2]{\AbstractNablaoid {\GradSym} {#1} {#2}}
\newcommand{\AbstractHess}[2]{\AbstractNablaoid {\HessSym} {#1} {#2}}
\newcommand{\AbstractHessInv}[2]{\AbstractNablaoid {\HessInvSym} {#1} {#2}}

\newcommand{\VoisLight}[1]{{\VoisSym #1}}
\newcommand{\Vois}[1]{{\VoisSym\left(#1\right)}}
\newcommand{\nVois}[2]{{\nVoisSym\left(#1,#2\right)}}

\newcommand{\Serie}[2]{ {\left \{ #1 \right \}}_{#2}}

\newcommand{\SerieElem}[2]{{#1}_{#2}}

\newcommand{\Component}[2]{{#1}^{#2}}

\newcommand{\Gradouze}[2]{\mathrm{Gradouze}}
\newcommand{\HessInvouze}[2]{\mathrm{Hessouze}}
\newcommand{\FunMAJouze}[2]{\mathrm{MAJouze}}

\newcommand{\FunDef}[5]{\begin{array}{lcrcl} 
#1 & : & #2 & \mapsto & #3 \\
 & & #4 & \rightarrow & #5
\end{array}
} 

\newcommand{\SpaceDim}[0]{m}
\newcommand{\FieldDim}[0]{n}
\newcommand{\LawDim}[0]{p}
\newcommand{\SetSpace}[0]{{\Reel^{\SpaceDim}}}
\newcommand{\SetField}[0]{{\Reel^{\FieldDim}}}
\newcommand{\SetLaw}[0]{{\Reel^{\LawDim}}}

\newcommand{\CanonFuncSym}{e}
\newcommand{\CanonFunc}[2]{\CanonFuncSym_{(#1,#2)}}
\newcommand{\CanonReelSym}{r}
\newcommand{\CanonBaseSym}{\CanonFuncSym}
\newcommand{\CanonBase}{\Serie\CanonBaseSym {(\Maille,i)}}
\newcommand{\LocCanonBase}[1]{\Serie \CanonBaseSym {#1}}
\newcommand{\kroneckerSym}{\delta}
\newcommand{\kronecker}[2]{\kroneckerSym_{#1 #2}}

\newcommand{\StepIdx}[0]{t}

\newcommand{\EVSym}[0]{V}
\newcommand{\EVat}[1]{{\EVSym}^{#1}}
\newcommand{\EVati}[0]{\EVat i}
\newcommand{\EVatip}[1]{\EVat {i+#1}}
\newcommand{\EVatim}[1]{\EVat {i-#1}}
\newcommand{\Time}[2]{#1_{#2}}
\newcommand{\Timet}[1]{\Time{#1}{\StepIdx}}
\newcommand{\Timetpu}[1]{\Time{#1}{\StepIdx+1}}
\newcommand{\ChargeSym}[0]{\rho}
\newcommand{\ChargeAt}[1]{{\ChargeSym^{#1}}}
\newcommand{\SpaceStep}[0]{d}
\newcommand{\waist}[0]{W}
\newcommand{\wave}[0]{A}
\newcommand{\ElecBaseSym}[0]{v}
\newcommand{\ElecBase}[1]{\ElecBaseSym_{#1}}

\newcommand{\ExampleN}[0]{N}
\newcommand{\ExampleStep}[0]{d}

\newcommand{\AffilMOPS}{\affiliation{Laboratoire Mat\'eriaux Optiques, Photonique et
Syst\`emes, 57070 Metz, France \\
 Unit\'e de Recherche Commune \`a l'Universit\'e Paul Verlaine --
Metz et Sup\'elec
				}
		}

\newcommand{\AffilIMS}{\affiliation{Information, Multimodality and Signal, Sup\'elec, 2, rue Edouard Belin, 57070 Metz, France
}}